\documentclass[journal]{IEEEtran}

\usepackage{color}
\usepackage[pdftex]{graphicx}
\usepackage{cite}
\usepackage{multirow}
\usepackage{graphicx}
\usepackage{booktabs}
\usepackage{mathtools}
\usepackage[none]{hyphenat}
\usepackage{url}

\graphicspath{{images/}}

\begin{document}

\title{LR-FHSS: Overview and Performance Analysis}

\author{
    Guillem~Boquet,
    Pere~Tuset-Peiró,~\IEEEmembership{Senior Member,~IEEE},
    Ferran~Adelantado,~\IEEEmembership{Senior Member,~IEEE},\\
    Thomas~Watteyne,~\IEEEmembership{Senior Member,~IEEE},
    Xavier~Vilajosana,~\IEEEmembership{Senior Member,~IEEE}
    \thanks{
        This research is co-financed by the EU Regional Development Fund within the ERDF Operational Program of Catalonia 2014-2020 with a grant of 50\% of total cost eligible, the SPOTS project (RTI2018-095438-A-I00) funded by the Spanish Ministry of Science, and the 2017 SGR 60 by the Generalitat de Catalunya.
    }
}

\markboth{This work has been submitted to IEEE Communications Magazine for peer review}{Boquet \MakeLowercase{\textit{et al.}}: LR-FHSS: Overview and Performance Analysis}

\maketitle

\begin{abstract}
Long Range-Frequency Hopping Spread Spectrum (LR-FHSS) is the new physical layer designed to address extremely long-range and large-scale communication scenarios, such as satellite IoT. At its core is a fast frequency hopping technique designed to offer higher network capacity while offering the same radio link budget as LoRa. Additionally, LR-FHSS finely manages packet transmission thanks to its design principles, enabling QoS policies on a per-packet basis. Given the notorious adoption of LoRaWAN in the IoT application landscape, this article is a reference for understanding how exactly LR-FHSS works, the performance it can offer, and its limitations and research opportunities.
\end{abstract}


\IEEEpeerreviewmaketitle

\section{Introduction}
\label{sec:intro}

The Internet of Things (IoT) relies on low-power wireless communication technologies and protocols to enable reliable wireless communication between distributed sensor and actuator devices. Over the last decade, we have seen the rise of Low-Power Wide Area Network (LPWAN) technologies~\cite{7815384}, with LoRaWAN becoming one of the most prominent players in the market thanks to long-range and robust communications, coupled with a simple network architecture that allows for solutions that are easy to deploy and manage. At the physical layer, LoRaWAN uses Long Range (LoRa), a robust Chirp Spread Spectrum (CSS) modulation developed by Cycleo, later acquired and commercialized by Semtech under the LoRa trademark. 
The LoRa physical layer is designed to prioritize uplink communication and ensure low power operation, limited by a low data rate (250~bps with spreading factor~12 and 125~kHz channel as the most restrictive) and kilometer-scale communication range. Different configurations of the physical layer are available providing different levels of chirp redundancy and thus trading off bandwidth utilization and robustness. LoRaWAN defines different bandwidth configurations per channel, ranging from 125~kHz to 500~kHz, depending on the regional parameters and available bandwidth. The access to these channels is based on pure ALOHA, limited by regional duty cycle regulations which constrain the time on air that can be utilized per device, further limiting the maximum achievable throughput \cite{adelantado17lorawan}.

\begin{figure}[h!]
    \centering
    \includegraphics[width=1.00\columnwidth]{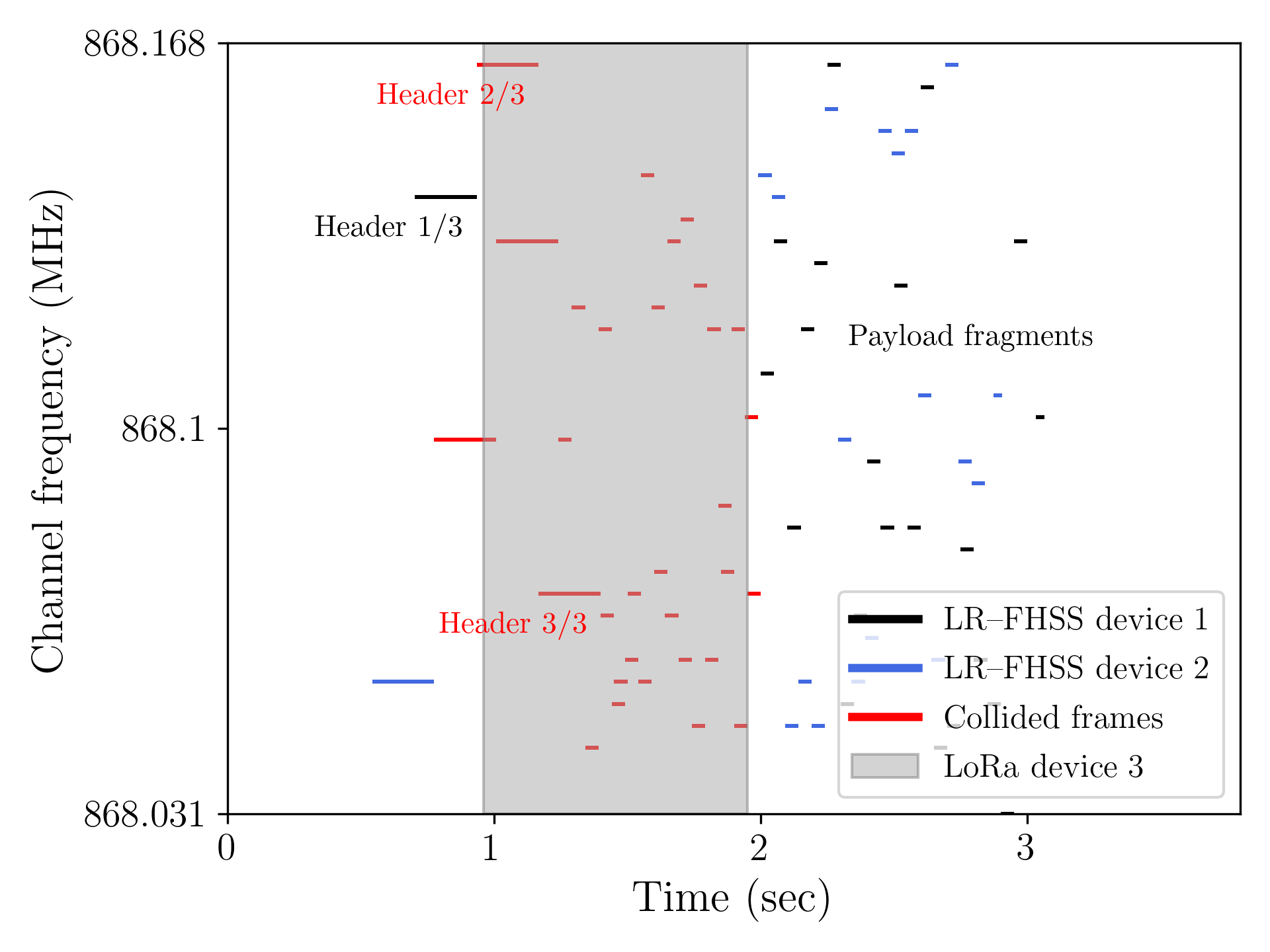}
    \caption{
        Two LR-FHSS (DR8) 30-B and one LoRa (DR0/SF12) 10-B packets transmitted simultaneously in the EU 868-870~MHz band (Channel~1). A packet transmission using LoRa occupies the whole channel bandwidth (125~kHz), whereas for LR-FHSS the fragments of a given packet are distributed over time in randomly selected subchannels (488~Hz) within the entire channel bandwidth (137~kHz). Despite several frame collisions, both LR-FHSS packets will be decoded successfully with high probability thanks to its redundancy. 
    }
    \label{fig:pack}
\end{figure}

The technology has been widely adopted. However, there are drawbacks and limitations to LoRaWAN, particularly in dense deployments where performance, that is, the overall network capacity, is severely limited by duty-cycle regulations and the use of simple Medium Access Control (MAC) protocols~\cite{10.1145/3293534}. In that sense, the research community has issued during the last years several proposals to address the fundamental operation of LoRaWAN at the physical and data-link layers~\cite{7954020, 8863372}, as well as the overall system operation~\cite{8430542}, using both analytic models and simulations~\cite{8090518}. In addition, the scalability and reliability issues of LoRaWAN have been the focus of recent research efforts towards ensuring fairness~\cite{reynders17power}, improving channel usage~\cite{8693695}, or even scheduling LoRa transmissions~\cite{8315103}.

As a novelty, Semtech has announced an extension of the LoRa physical layer called LR-FHSS, which stands for Long Range-Frequency Hopping Spread Spectrum~\cite{LoRa-E-alliance}. The extension is motivated by emerging use cases with increasingly larger and denser network deployments, including satellite-scale LoRaWAN networks. Its main goal is to increase network capacity and robustness by adopting the basis of FHSS modulation technique, while keeping the same communication range than LoRa and meeting the European Telecommunications Standards Institute (ETSI)~\cite{etsi2019electromagnetic}, the Federal Communications Commission (FCC)~\cite{fcc} and the Association of Radio Industries and Businesses (ARIB)~\cite{aribt108} regulations. Additionally, LR-FHSS has been designed to introduce higher levels of network flexibility, targeting applications that require differentiated service levels. 

We expect LR-FHSS to have a big impact on LPWAN and enable viable satellite IoT solutions. Therefore, in this article, we provide an overview of the technology and its performance, comparing it to today's LoRa, and validate the scalability gain announced by Semtech. We also identify open research issues and directions for this new physical layer.

The remainder of this article is organized as follows. Section~\ref{sec:operation} presents an overview of LR-FHSS. Section~\ref{sec:performance} studies the scalability of LR-FHSS networks and compares it to LoRa networks. Section~\ref{sec:research} outlines open research questions for LR-FHSS networks. Finally, Section~\ref{sec:conclusions} concludes the article.

\section{LR-FHSS Operation}
\label{sec:operation}
LoRaWAN prioritizes uplink capacity and limits downlink transmissions to sporadic data or control packets. LR-FHSS is a fast FHSS modulation used for uplink only; downlink communication is achieved with current LoRa, since the same radios can switch between modulations.  

LR-FHSS relies on two bit rates (162 and 325~bps), according with the new LoRaWAN Data Rate (DR) modes (Section \ref{sec:dr}).
To initiate transmission of a packet (Section \ref{sec:header}), end-devices randomly select one of the LR-FHSS channels available and use a pure ALOHA access mechanism.
As illustrated in Fig.~\ref{fig:pack}, the channels are divided into several subchannels (Section \ref{sec:ch}), which the transmitter uses to change the carrier frequency according to a certain hopping pattern at each transmission (Section \ref{sec:fhop}).
First, several replicas of the header are transmitted. The number of replicas is defined by each LR-FHSS DR. The packet payload is then split into fragments with a duration of $\sim$50~ms. Contrary to the packet header, only a single copy of each fragment is transmitted. All headers and payload fragments are sent consecutively on each subchannel determined by the frequency hopping sequence of the transmitter. 
At the other end, the gateway reassembles the packet payload using the information within a header.

Unlike for LoRa channels, as long as LR-FHSS transmissions fall entirely in the gateway processed bandwidth, the packets can be demodulated.
This means that the gateway does not need to know in advance any channel hopping sequence, nor the exact channels frequencies and bandwidths.
Therefore, different devices can use a different spreading bandwidth and may not all have the channel at the same frequency.
It also implies that multiple transmitters can operate at the same time, provided they use different channel hopping sequences and the gateway is able to listen to the whole channel bandwidth at the same time.
This increases the complexity of signal detection at the receiver, compared to LoRa. However, it allows hundreds of packets to be received simultaneously, which is appropriate for satellite-scale networks where the number of devices interfering within the coverage area of a gateway located at space is much greater than the number of devices found in current LoRaWAN use cases.

LR-FHSS can be used with limited support (no intrapacket hopping) for current SX1272/76 devices through a firmware update. 
In contrast, the newer SX1261, SX1262 and SX1268 modems are fully LR-FHSS compatible and can take advantage of its FHSS modulation. 
Lastly, demodulation of LR-FHSS signals can be achieved by all gateways using the SX1302 modem with a firmware upgrade.

\subsection{Data Rates}
\label{sec:dr}

LR-FHSS (LR-FHSS) DR modes are shown in Table~\ref{tab:loraparams}. In the EU863-870 band, those are DR8/10 (slower, higher robustness) and DR9/11 (faster, lower robustness).
DR8 and DR10 use a coding rate of 1/3, 3~header repetitions, and a physical bit rate of 162~bps. DR9 and DR11 use a coding rate of 2/3, 2~header repetitions, and a physical bit rate of 325~bps. The aforementioned coding rates are used to convolutionally encode the payload bits that are decoded using a Viterbi decoder. Although the specification allows for header repetitions ranging from 1 to 4, only the two configurations above are used.
This redundancy is key to ensuring the robustness of LR-FHSS when fragments are spread in frequency. For example, the gateway device is able to reassemble a packet transmitted using DR9 with high probability, even if 1 of the 2 headers and 1/3 of the bits within the payload fragments are lost. These mechanisms mitigate the impact of interference from other LoRa and LR-FHSS devices operating in the same band, as well as devices using other wireless technologies.

\begin{table*}[ht!]
    \centering
    \caption{
        LR-FHSS main specifications and parameters for EU and US regions.
    }
    \begin{tabular}{|l|c|c|c|c|c|c|}
    \hline
    \textbf{Region} & \multicolumn{4}{c|}{\textbf{European Union (ETSI, 863-870 MHz)}} & \multicolumn{2}{c|}{\textbf{United States (FCC, 902-928 MHz)}}
    \\ \hline
    LoRaWAN data rate alias & \multicolumn{1}{c|}{DR8} & \multicolumn{1}{c|}{DR9} & \multicolumn{1}{c|}{DR10} & \multicolumn{1}{c|}{DR11} & \multicolumn{1}{c|}{DR5} & \multicolumn{1}{c|}{DR6}
    \\ \hline
    LR-FHSS number of channels & \multicolumn{1}{c|}{7} & \multicolumn{1}{c|}{4} & \multicolumn{1}{c|}{7} & \multicolumn{1}{c|}{4} & \multicolumn{2}{c|}{8}
    \\ \hline
    LR-FHSS OCW (kHz) & \multicolumn{2}{c|}{137} & \multicolumn{2}{c|}{336} & \multicolumn{2}{c|}{1523}
    \\ \hline
    LR-FHSS OBW (Hz) & \multicolumn{6}{c|}{488}
    \\ \hline
    \begin{tabular}[c]{@{}l@{}}Minimum separation between \\ LR-FHSS hopping carriers (kHz) \end{tabular} & \multicolumn{4}{c|}{3.9} & \multicolumn{2}{c|}{25.4}
    \\ \hline

    \begin{tabular}[c]{@{}l@{}}Number of physical carriers available \\ for frequency hopping in each OCW channel\end{tabular} & \multicolumn{2}{c|}{280 (8x35)} & \multicolumn{2}{c|}{688 (8x86)} & \multicolumn{2}{c|}{3120  (52x60)}
    \\ \hline
    \begin{tabular}[c]{@{}l@{}}Number of physical carriers usable for \\ frequency hopping per end-device transmission\end{tabular} & \multicolumn{2}{c|}{35} & \multicolumn{2}{c|}{86} & \multicolumn{2}{c|}{60}
    \\ \hline
    Coding rate & 1/3 & 2/3 & 1/3 & 2/3 & 1/3 & 2/3
    \\ \hline
    Physical bit rate (bits/s) & 162 & 325 & 162 & 325 & 162 & 325
    \\ \hline
    Max. MAC payload size (bytes)& 58 & 123 & 58 & 123 & 125 & 125
    \\ \hline
    Max. MAC payload fragments & 61 & 64 & 61 & 64 & 130 & 65
    \\ \hline
    Header replicas & 3 & 2 & 3 & 2 & 3 & 2
    \\ \hline
    PHY header duration per replica (seconds) & 0.233 & 0.233 & 0.233 & 0.233 & 0.233 & 0.233
    \\ \hline
    PHY time on air (seconds) & 0.70 + 3.06 & 0.47 + 3.19 & 0.70 + 3.06 & 0.47 + 3.19 & 0.70 + 6.48 & 0.47 + 3.24
    \\ \hline
    \end{tabular}
    \label{tab:loraparams}
\end{table*}

\subsection{Packet Format}
\label{sec:header}
\begin{figure*}[h!]
    \centering
    \includegraphics[width=1.30\columnwidth]{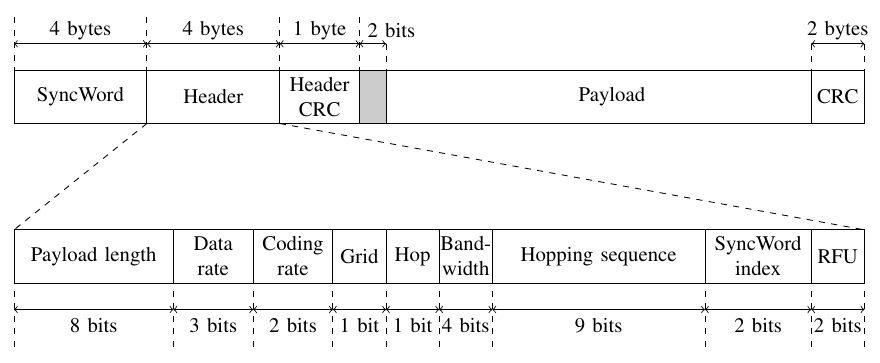}
    \caption{
        LR-FHSS packet structure. 
        While the header is 5~bytes long (including header-CRC), the radio transmits a total of 80~bits as it uses a convolutional code with a 1/2 code rate.
        Between the header and the payload, the transceiver waits a time equal to 2~bits (shaded gray) to allow the receiver to decode and process the header.
        The header is 114~bits long, the payload can be up to 125~bytes.
    }
    \label{fig:PHYpack}
\end{figure*}

The LR-FHSS packet, Fig.~\ref{fig:PHYpack}, is composed of a SyncWord, a PHY Header (including CRC) and the payload (including CRC). The first transmitted fragments (2 or 3 replicas) contain the entire header (SyncWord, Header, Header CRC). The header is transmitted at a fixed bit rate for a duration of 0.233~s, containing information about channel hopping sequence, payload length, data rate, number of header replicas and coding rate. Each packet header contains the necessary information such that the gateway can compute the exact list of frequencies that the packet will be using. Thus, the gateway must receive at least one copy of the header to be able to detect and reassemble the packet.
Note in Fig.~\ref{fig:pack} the transmission of 3 consecutive header replicas with the same duration, which are not split across multiple fragments despite being longer than $\sim$50~ms. The next fragments only contain the payload and are transmitted at the configured data rate. Recall that its information is encoded in such a way that even if 1/3 of the fragments are lost, the information can still be recovered and the packet can be reassembled with high probability.

Additionally, LR-FHSS enables to select the modulation, number of header repetitions, coding rate and frequency spreading bandwidth, on a packet per packet basis. This allows a fine grained network resource management based on the link quality that allows to reach the required QoS for each device.

\subsection{Channels and subcarriers}
\label{sec:ch}

Similar to LoRaWAN, the frequency band is split into different Operating Channel Width (OCW) channels. The number of channels available depends on the region of the world in which the network operates, thus a white-list is used to mark those channels. In North America (FCC 902-928~MHz), the specification defines 8 LR-FHSS OCW channels with 1.523~MHz bandwidth and center frequencies equal to $(903+1.6\, n)$~MHz, $n=0,\dots,7$. In Europe (ETSI 863-870~MHz), the LR-FHSS OCW channels are defined to have a bandwidth equal to 137~kHz or 336~kHz depending on the selected DR. The number of channels that can be used depends on the specific gateway technology used by the network operator. The number of LR-FHSS channels supported by a gateway depends on the number of Digital Signal Processors (DSPs) used, each DSP covering 1.523~MHz. For example, in a gateway with one DSP and 200~kHz channel spacing, end-devices can use up to 7 LR-FHSS channels when using a 137~kHz OCW channel for the uplink. 

Each LR-FHSS OCW channel is divided into several Occupied Band Width (OBW) physical channel subcarriers with a bandwidth of 488~Hz. For example, the 137~kHz channel bandwidth in the EU 863-870~MHz regional parameters enables 280 LR-FHSS OBW subcarriers in a single LR-FHSS OCW channel, therefore supporting simultaneous transmission. Because the regional regulations impose restrictions on the time and bandwidth occupation, subcarrier hopping policies must have a minimum frequency hop. That is, two consecutive subcarriers of a hopping sequence must be separated by a minimum distance. In the EU band, the minimum frequency shift is 3.9~kHz for each fragment~\cite{etsi2019electromagnetic}, thus creating 8 simultaneous grids of 35 usable subcarriers in each of the 137~kHz LR-FHSS OCW channels for DR8/9. In the US band, the minimum separation between LoRa–E hopping subcarriers is 25.4~kHz. Similarly, this results in 52~simultaneous groups of 60 usable subcarriers for each particular 1.523~MHz OCW channel.

Table~\ref{tab:loraparams} summarizes the LR-FHSS physical layer both for ETSI and FCC regulations. In some configurations, the relatively small number of subcarriers in a group (35 subcarriers in each of the 8 groups in an EU 137~kHz OCW channel) may cause fragment collisions, as these are sent in a pseudo-random frequency hopping pattern. However, the probability of concurrent transmissions, that is, two end devices selecting the same frequency hopping pattern at approximately the same time, is very low.

\subsection{Frequency Hopping Policy}
\label{sec:fhop}

For each uplink packet, the device randomly selects a channel amongst the ones enabled that can support the data rate. The transmission starts on a random physical subchannel (grid) inside the LR-FHSS spreading bandwidth, then follows a pseudo-random frequency hopping pattern computed by the device. Given a specific OCW frequency offset, the hopping pattern is obtained as the outcome of a simple 32-bit hash function executed by the LR-FHSS device, modulo the number of available channels in the grid or physical subcarriers usable for channel hopping per end-device transmission (35 or 86 in EU and 60 in USA, Table~\ref{tab:loraparams}). In particular, the input of the hash function is determined by the result of a device-specific 9-bit random number (header's hopping sequence field in Fig.~\ref{fig:PHYpack}) plus the product of the current fragment number and $2^{16}$. 
Finally, the hash is multiplied by the minimum frequency separation between LR-FHSS hopping carriers to comply with the regulations of the transmission band.
All of this ensures a pseudo-random pattern where all physical channels are statistically used equally.

\section{Key Performance Aspects}
\label{sec:performance}

Understanding the performance of LR-FHSS under saturation conditions and how it compares to LoRa are essential questions to any LR-FHSS prospective user. In this section, we evaluate and compare its performance from the end-device and the network scalability perspective. The software used in this article is publicly available at \url{https://github.com/wine-uoc/}.

To evaluate its key performance aspects we developed a packet-based network simulator that implements the basic functionality of both protocols. The simulator uses an exponential distribution to model the random time between independent packet arrivals, where the arrival rate is established at the maximum transmission rate allowed by the duty cycle regulations. Packets are allocated in a discrete time and frequency space with a millisecond time resolution and OBW subcarrier granularity. Simultaneous transmissions only cause a collision if they both select the same OBW subcarrier and overlap in time. For each LR-FHSS device, a channel hopping sequence is created at the beginning, according to the procedure described in Section~\ref{sec:operation} and the parameters presented in Table~\ref{tab:loraparams}. That is, packets are divided into several fragments and spread following the device's hopping sequence at transmission time. At reception, a LR-FHSS packet is successfully decoded if minimum one of the headers and 1/3 (or 2/3) of the payload have not collided. For LoRa, packets are transmitted using the entire available channel bandwidth, and decoded only if they have not collided. 

We performed extensive simulations assuming ideal channel conditions and the EU 863-870~MHz band, which imposes the most restrictive per-channel duty cycle of 1\%. Only non-colliding packets are successfully received, hence, the results presented should be considered as a lower performance bound of both LoRa and LR-FHSS. 

\subsection{End-device Capacity}

\begin{figure}[]
    \centering
    \includegraphics[width=1.00\columnwidth]{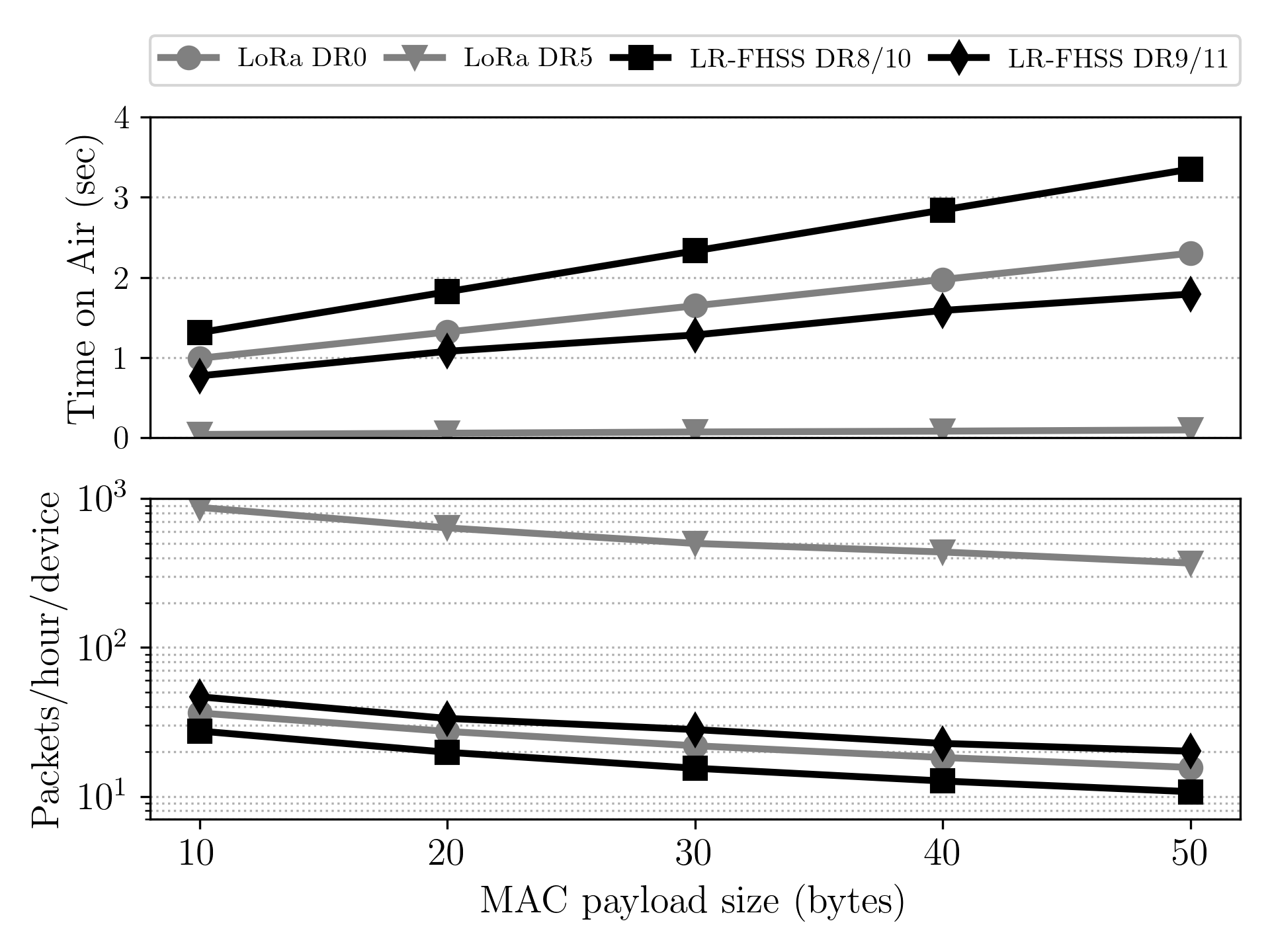}
    \caption{
        Relationship between the packet duration (seconds) and the maximum per end-device transmission rate (packets/hour) when abiding to the EU 868-870~MHz band regulation, which imposes a 1\% duty cycle per device and channel.
    }
    \label{fig:xTOALMBD}
\end{figure}

Fig.~\ref{fig:xTOALMBD} shows the time on air \textit{(top)} and the number of packets per hour per end-device \textit{(bottom)} for different MAC payload sizes (10 and 50~B) for LoRa DR0 (SF12), DR5 (SF7 125~kHz), and the LR-FHSS data rates available in Europe (DR8/10 and DR9/11). As depicted, LoRa DR5 provides the highest number of packets per hour per end-device (369.1 to 873.5), whereas LR-FHSS DR8/10 provides the least (10.7 to 27.5). In contrast, LoRa DR0 provides between 15.6 and 36.3 packets per hour per end-device, whereas LR-FHSS DR9/11 provides similar capacity regardless of the MAC payload size, having between 20.1 and 46.6~packets per hour per end-device.

As expected, the lower LR-FHSS DR combined with the imposed duty cycle regulations limit the number of packets that can be transmitted. Observing these results, one could conclude that it does not provide any real benefit in terms of individual end-device capacity. However, LR-FHSS is not designed to increase the number of packets that can be transmitted by each end-device independently. Rather, its strength lays in the overall network capacity increase provided by the statistical multiplexing of combining time and frequency diversity.

\subsection{Network Capacity}
\begin{figure}[]
    \centering
    \includegraphics[width=1.00\columnwidth]{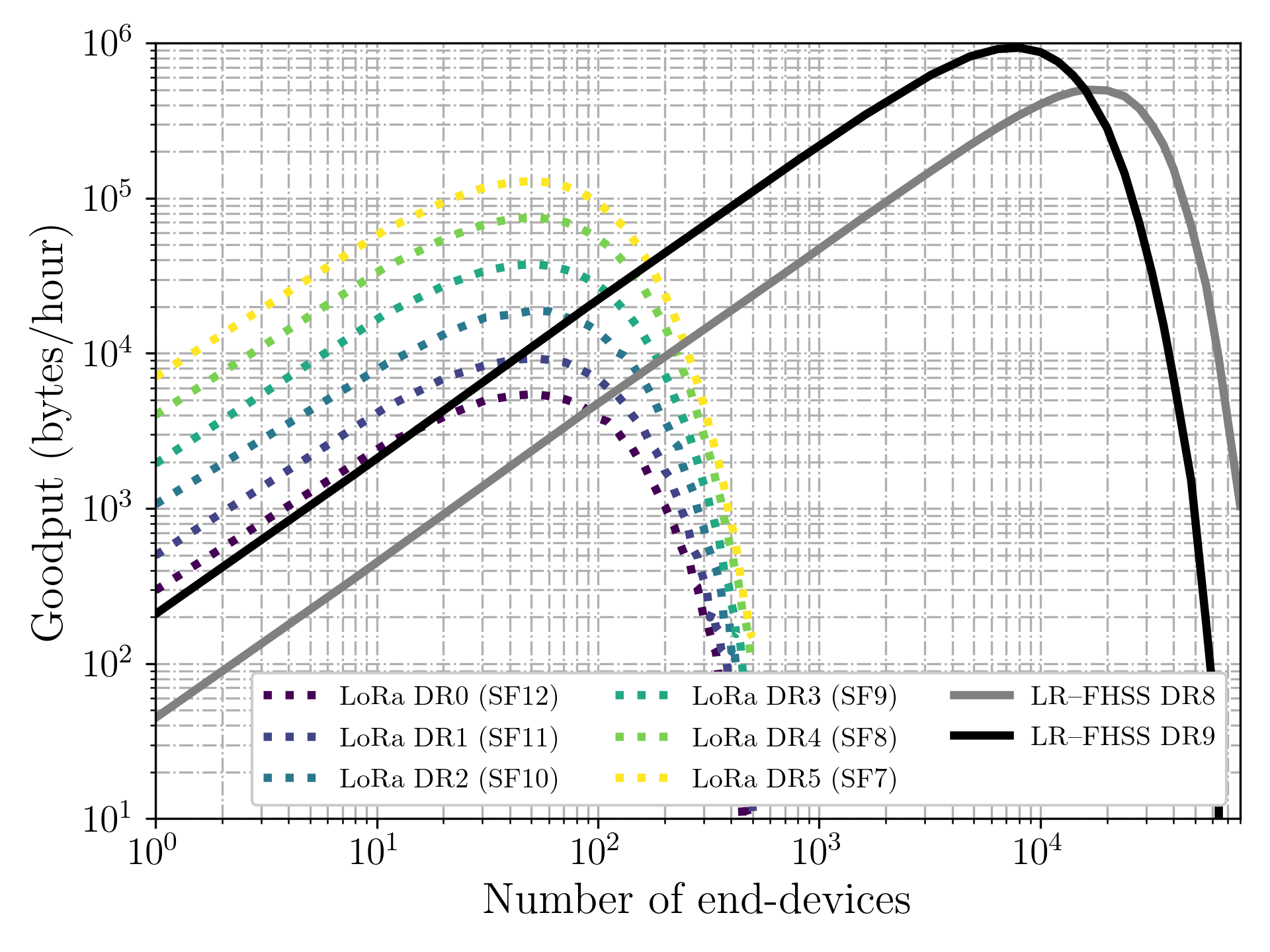}
    \caption{
        Total useful bytes received per hour (goodput) when a given number of LoRa and LR-FHSS end-devices transmit 10~B of payload at the maximum 1\% duty cycle allowed by the EU 868-870~MHz regulation.
        Both LoRa and LR-FHSS results are obtained considering 125~kHz and  137~kHz channels, respectively.
    }
    \label{fig:RXdev}
\end{figure}

Fig.~\ref{fig:RXdev} compares a 125~kHz LoRa channel with a 137~kHz LR-FHSS channel (supporting 8 grids of 35 physical subcarriers), and shows how LR-FHSS scales with the number of end-devices transmitting at the maximum possible rate using a payload of 10~B. The number of supported end-devices peaks at 50 when using LoRa regardless of the DR used.
This can be explained by the fact that the network is limited at the data-link layer by the ALOHA MAC protocol, and at the physical layer by the radio duty cycle regulation (1\% at the EU 868-870~MHz band). Higher DRs provide higher goodput because each DR (almost) doubles the data rate of the previous one, hence the transmission time on air of the packet decreases proportionally and more packets (information) can be fitted, offering the same channel load. In other words, the total offered load remains constant despite the DR used because devices' maximum transmission rate is fixed by the duty-cycle regulation. A higher DR provides additional capacity, resulting in an unchanged collision probability in this scenario. Simulations performed with payload sizes up to 50~B showed the same behavior.

In contrast, the number of end-devices that can transmit simultaneously in LR-FHSS increases, leading to a significant increase in the goodput of the entire network as it scales. Notably, LR-FHSS ensures a goodput proportional to the throughput that is one order of magnitude larger than LoRa for the fastest of the DRs (DR5) and two orders of magnitude for the slowest (DR0). The maximum goodput while using DR9/8 and a payload size of 10~B occurs when 8,000/18,000~end-devices transmit at the maximum allowed rate (given the payload size, Fig.~\ref{fig:xTOALMBD}) or, in other words, 370,000/500,000~packets with 10-B payload are generated per hour within the network. Above that number of end-devices or packet generation, the performance of DR9 decreases faster than DR8, causing DR8 to scale better. This is because DR8 repeats the header 3~times (2~times in DR9) and transmits more redundant information for error correction, which increases robustness against collisions.
Thus, we can extrapolate that DR8 performs better than DR9 in terms of goodput on channels with significant interference. For example, in Fig.~\ref{fig:RXdev} when 16,000 or more end-devices are transmitting simultaneously at the maximum allowed rate, the LR-FHSS DR8 network outperforms DR9.

\begin{figure}[]
    \centering
    \includegraphics[width=1.00\columnwidth]{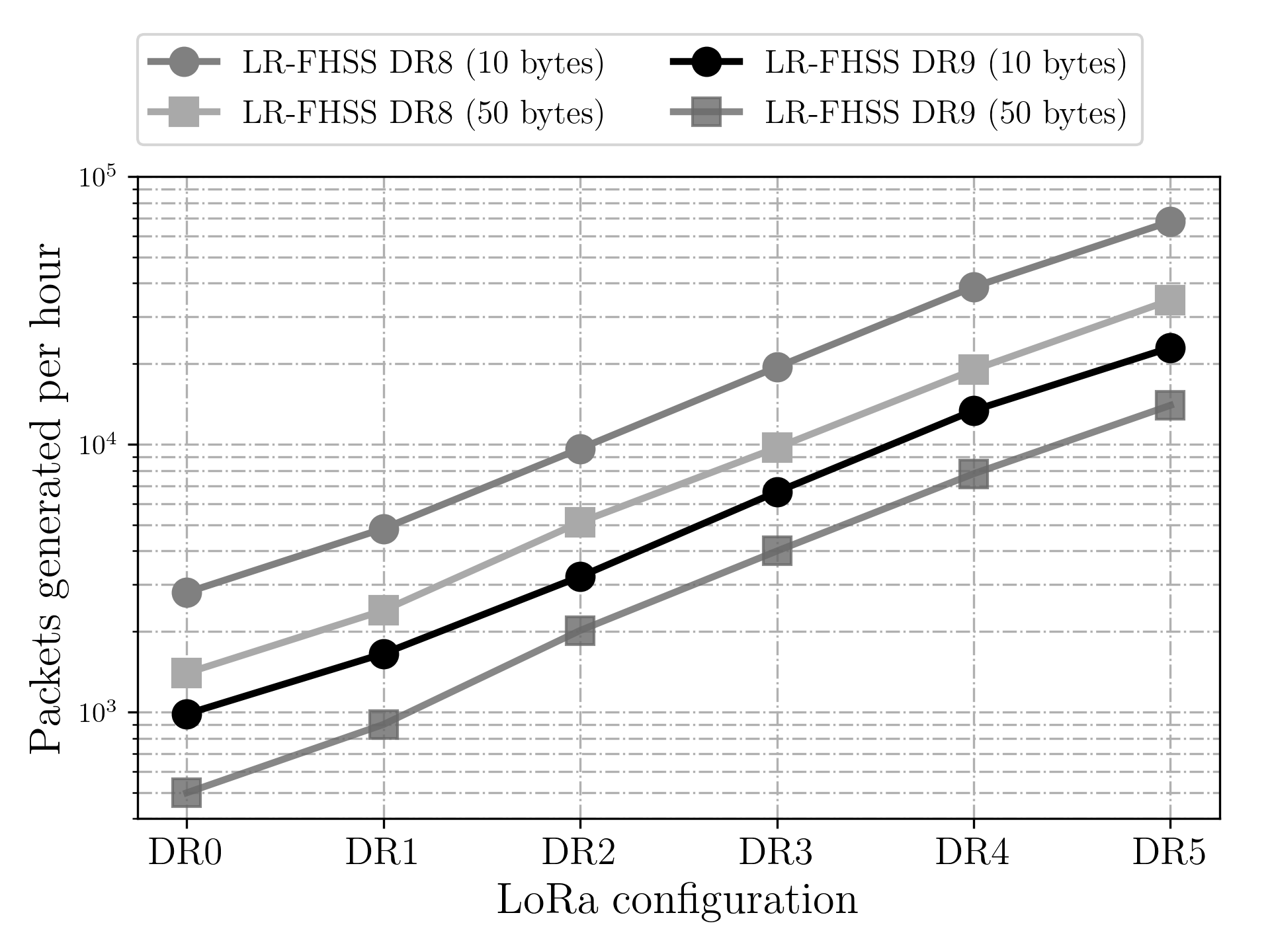}
    \caption{
        Comparison between LoRa and LR-FHSS DRs performance.
        Markers show the crossover points in number of generated packets per hour required for LR-FHSS to provide a better goodput than LoRa, as a function of payload size.
    }
    \label{fig:iPOINTS}
\end{figure}

Fig.~\ref{fig:iPOINTS} shows the crossover points between LoRa and LR-FHSS goodput as a function of the number of generated packets per hour for the different LoRa DR configurations and MAC payload sizes (10 and 50~B). If we compare LR-FHSS with devices transmitting 10~B of payload size against the lowest LoRa data rate configuration (DR0), the LR-FHSS network requires a network load greater than 2,800 and 985~packets/hour for DR8 and DR9, respectively, to provide a higher goodput than the LoRa DR0 network. At the other end, LR-FHSS requires around 68,000~packets/hour for DR8 and 23,000~packets/hour for DR9 to provide higher goodput compared to the LoRa DR5 configuration. This is an exponential growth with a growth rate of almost~2, similar to the increase in the available data rate when the LoRa DR is increased from DR0 to DR5. The same growth rate is found when devices transmit 50~B of payload size.  However, the crossover points occur for lower values of generated packets per hour, meaning that LR-FHSS is more efficient in terms of goodput when the ratio between payload and overhead increases. Interestingly, the required number of packets/hour needed for LR-FHSS to be greater than LoRa is divided by 2 if devices increase the payload size from 10 to 50-B. For example, regarding LR-FHSS DR8 and LoRa DR0, the crossover point for 10~B is at 2,800~packets/hour while for 50~B is at 1,400~packets/hour.

Finally, notice that the results presented are for a single LoRa and LR-FHSS channel and, hence, the overall network capacity has to be multiplied by the number of available channels and DR configurations in the band of interest. For example, transmitting a 10-B payload in the EU 868-870~MHz band, a legacy LoRa DR0 network could ideally support a total load of 96,000~packets/hour at its peak efficiency, when considering the use of 8~independent 125~kHz channels and 6 DRs simultaneously. In contrast, the total capacity of a EU LR-FHSS network would be around 3.5M~packets/hour when using DR8, and 1.48M~packets/hour when using DR9. This assumes 7 and 4 channels with 8~frequency grids each, as summarized in Table~\ref{tab:loraparams}, which represents a 36$\times$ and 15$\times$ capacity increase for LR-FHSS when compared to LoRa.

\subsection{Use Cases}

With a lower data rate than DR0 (SF12), the time on air for a LR-FHSS packet and the number of packets that an end-device can transmit during a period of time is smaller than LoRa.
Yet, this nominal under-performance does not translate into an effective reduction of the goodput because fragment spreading reduces collision probability. This mode of operation can also be exploited to scale the network in terms of the number of end-devices supported concurrently. That is, LR-FHSS slices the spectrum so that multiple end-devices can use the same channel with a insignificantly low collision probability. Hence, LR-FHSS offers higher network scalability when compared to LoRa DR0. Considering the presented results, we can conclude that LR-FHSS provides better network scalability than LoRa thanks to the adoption of FHSS modulation technique. In particular, the 162~bps LR-FHSS DRs provide two orders of magnitude more network capacity, while offering the same radio link budget than LoRa DR0 (SF12). In addition, LR-FHSS retains the long-range communication characteristics of LoRa.
This makes LR-FHSS suitable for terrestrial and satellite deployments, where a high end-device density or a large gateway coverage area leads to an increased level of interference. Summarizing, LR-FHSS allows the necessary $>$155~dB link margin for low Earth orbit (LEO) satellite IoT plus the capacity to receive hundreds of packets simultaneously.

\section{Open Research Challenges}
\label{sec:research}
This section discusses two main open research challenges presented by LR-FHSS: optimal selection of frequency hopping sequences and coexistence with legacy LoRa networks.

\subsection{Optimal Frequency Hopping Sequences}
The LR-FHSS specification uses a simple frequency hopping pattern based on a 32-bit hash function. These hopping sequences create grids on the OCW channels that space subcarriers a minimum frequency distance according to the region's specification. While these schemes ensure a spreading of the packets among the subcarriers, optimal policies can be designed to address specific channel conditions. In particular, schemes that dynamically adapt sequences to continuously match a changing environment. Because LR-FHSS networks can manage the spectrum used by each device individually, adaptive frequency hopping~\cite{popovski2006strategies} can either aim to maintain data throughput or allow flexible fine-grained network resource management to achieve the desired QoS per device. Hence, an important research topic is to evaluate different hopping sequences to maximize the properties of the FHSS scheme. Also, it would be interesting to explore sequence management policies that determine how and when these dynamic sequences need to be updated.

\subsection{Coexistence with Legacy LoRa Networks}

As LoRaWAN deployments will integrate end-devices using different LoRa DRs, including the novel LR-FHSS, and will have to coexist with other LoRaWAN networks using different combinations of DRs, coexistence is a key aspect to investigate. In particular, the results evaluating the coexistence between LoRa and LR-FHSS DRs will allow to build recommendations and best-practice strategies for deploying these technologies and ensuring their performance. Such recommendations may be to dedicate some channels to LR-FHSS, or to limit the proportion of different DRs in each channel. Another topic related to coexistence is adaptive data rate (ADR), for which the role of LR-FHSS needs to be defined. In that regard, LoRaWAN may have to include new functionalities to adapt to an increase in network load by dynamically switching to LR-FHSS when the goodput degrades.

\section{Conclusions}
\label{sec:conclusions}

LR-FHSS is a physical layer designed to offer flexibility for differentiated services and increase the scalability of LoRaWAN networks, retaining its long-range communication  characteristics. This article has presented LR-FHSS, evaluated its performance and highlighted its limitations. Results showed a significant increase in network scalability at the cost of individual end-device capacity. This article has also discussed open research challenges including mechanisms for optimizing frequency hopping sequences, policies for correctly planning and adapting the network to load, as well as LoRaWAN networks coexistence.

\bibliographystyle{IEEEtran}
\bibliography{references}

\begin{IEEEbiographynophoto}{Guillem Boquet}
received his M.Sc. and PhD. in Telecommunications Engineering from Universitat Aut\`onoma de Barcelona (UAB) in 2014 and 2020, respectively. He is currently a Researcher at the Wireless Networks (WiNe) group of the Universitat Oberta de Catalunya (UOC).
\end{IEEEbiographynophoto}

\begin{IEEEbiographynophoto}{Pere Tuset-Peiró}
(M'12, SM'18) is Assistant Professor at the Universitat Oberta de Catalunya (UOC) and Senior Researcher at the Wireless Networks (WiNe) group. He received his M.Sc. in Telecommunications Engineering from Universitat Politècnica de Catalunya (UPC), and his Ph.D. in Network and Information Technologies from Universitat Oberta de Catalunya (UOC).
\end{IEEEbiographynophoto}

\begin{IEEEbiographynophoto}{Ferran Adelantado} 
(M'08, SM'19) is Associate Professor at the Universitat Oberta de Catalunya (UOC) and Senior Researcher at the Wireless Networks (WiNe) group. He holds a M.Sc. degree in Telecommunications Engineering (2001) and a PhD (2007) from the Universitat Politècnica de Catalunya (UPC).
\end{IEEEbiographynophoto}

\begin{IEEEbiographynophoto}{Thomas Watteyne}
(sM'06, M'09, SM'15) holds a PhD in Computer Science (2008), an MSc in Networking (2005) and an MEng in Telecommunications (2005) from INSA Lyon, France. He is a Research Director at Inria in Paris, leading the EVA research team, and network designer at Analog Devices. He previously worked at Orange Labs and UC Berkeley.
\end{IEEEbiographynophoto}

\begin{IEEEbiographynophoto}{Xavier Vilajosana}
(M'09, SM'15) received his B.Sc. and M.Sc in Computer Science from Universitat Politècnica de Catalunya (UPC) and his Ph.D. in Computer Science from the Universitat Oberta de Catalunya (UOC). He has been a researcher at Orange Labs, HP and UC Berkeley. He is now Professor at UOC.
\end{IEEEbiographynophoto}

\end{document}